\documentclass[conference]{IEEEtran}
\ifCLASSINFOpdf
\else
\fi
\usepackage{graphicx}
\usepackage{makecell}

\begin{document}
%
\title{Machine learning on Crays to optimise petrophysical workflows in oil and gas exploration}

\author{\IEEEauthorblockN{Nick Brown}
\IEEEauthorblockA{EPCC\\University of Edinburgh\\
Bayes Building, Edinburgh}\\
\IEEEauthorblockN{Lucy MacGregor}
\IEEEauthorblockA{Cognitive Geology\\Torphichen Street\\
Edinburgh}
\and
\IEEEauthorblockN{Anna Roubickova}
\IEEEauthorblockA{EPCC\\University of Edinburgh\\
Bayes Building, Edinburgh}\\
\IEEEauthorblockN{Michelle Ellis}
\IEEEauthorblockA{Rock Solid Images\\Houston, Texas}
\and
\IEEEauthorblockN{Ioanna Lampaki}
\IEEEauthorblockA{EPCC\\University of Edinburgh\\
Bayes Building, Edinburgh}\\
\IEEEauthorblockN{Paola Vera de Newton}
\IEEEauthorblockA{Rock Solid Images\\Houston, Texas}

}


%


\maketitle

\begin{abstract}
The oil and gas industry is awash with sub-surface data, which is used to characterize the rock and fluid properties beneath the seabed. This in turn drives commercial decision making and exploration, but the industry currently relies upon highly manual workflows when processing data. A key question is whether this can be improved using machine learning to complement the activities of petrophysicists searching for hydrocarbons. In this paper we present work done, in collaboration with Rock Solid Images (RSI), using supervised machine learning on a Cray XC30 to train models that streamline the manual data interpretation process. With a general aim of decreasing the petrophysical interpretation time down from over 7 days to 7 minutes, in this paper we describe the use of mathematical models that have been trained using raw well log data, for completing each of the four stages of a petrophysical interpretation workflow, along with initial data cleaning. We explore how the predictions from these models compare against the interpretations of human petrophysicists, along with numerous options and techniques that were used to optimise the prediction of our models. The power provided by modern supercomputers such as Cray machines is crucial here, but some popular machine learning framework are unable to take full advantage of modern HPC machines. As such we will also explore the suitability of the machine learning tools we have used, and describe steps we took to work round their limitations. The result of this work is the ability, for the first time, to use machine learning for the entire petrophysical workflow. Whilst there are numerous challenges, limitations and caveats, we demonstrate that machine learning has an important role to play in the processing of sub-surface data.
\end{abstract}


%
\IEEEpeerreviewmaketitle

\section{Introduction}
The oil and gas industry is awash with sub-surface data, which is used to characterize the rock and fluid properties beneath the seabed. This information in turn drives commercial decision making, exploration and exploitation planning. However the business and technology models employed in upstream geology and geophysics have scarcely changed since the 1980s and are entirely unsuitable for the modern digital world. As such, the wealth of available data is currently poorly utilized and the full value seldom realized. Making better use of information, using modern data analytics techniques, and presenting this information in a way that is immediately useful to geologists and decision makers has the potential to dramatically reduce time to decision and the quality of the decisions that are made.  

In this paper we concentrate on one aspect of the problem, streamlining petrophysical workflows \cite{petro-workflow}. In such workflows well log data is used to quantitatively characterise the rock, providing a ground truth from which rock physics relationships can be constructed, and providing calibration between measurable geophysical properties and the underlying rock and fluid properties of interest. Using such relationships, geophysical attributes can be used to determine properties such as the porosity, total Clay volume or fluid saturation. Examples of manual use of well log data in this context are provided by \cite{welllog1} and \cite{welllog2}.

Well log data itself is collected from drilled boreholes, where numerous physical measurements are collected are collected downhole. Raw data is collected and then manually interpreted, via a petrophysical workflow, into processed log suites containing mineralogy, lithology and fluid content of the sub-surface. In a regional context, well log databases provide valuable insights into the variations in rock and fluid properties of the sub-surface and underlying control factors, which can be used to better understand existing acreage and prospects, along with exploring new areas.  For example previous work in \cite{welllog3} and \cite{welllog4} demonstrated the application of a regional rock physics relationships to understand electrical anisotropy in the Barents Sea and Malay basin respectively.

In order to be useful in such an analysis, raw well log data must be conditioned, erroneous data corrected and missing data estimated.  Mineralogy, lithology, porosity and fluid properties must be determined and from this rock physics models can then be constructed. This can be lengthy process, typically carried out by specialist petrophysicists. A number of the steps in a petrophysical workflow, if distilled down to their fundamentals, are pattern recognition problems: we have a known set of input curves (usually physical measurements of the earth such as gamma ray, neutron porosity, density, and resistivity among others), and we want to predict a series of output curves (for example porosity, clay content, and fluid saturation) based on the characteristics of the measurements. In this paper we present an investigation into the use of supervised machine learning approaches to dramatically streamline this petrophysical workflow.

Supervised machine learning approaches rely on labelled training data, from which relationships between the predictor variables, and the target variables are built. In general the use of these techniques in the oil and gas industry \cite{mlfluidsaturation}\cite{electrical_paper} is still very much in its infancy. Previous work has demonstrated some success around using supervised learning for one part of the workflow, for instance \cite{newcastle_paper} studied the prediction of the fractional composition of two minerals, which is part of the overall mineralogy composition, but crucially these studies only include a small set of wells to train and test the model from. Furthermore, the quality of data in these wells is uncertain which is largely driven by a lack of access to high quality wider ranging data sets. Crucially, such a small set of wells tends to make the problem much simpler as these often represent wells close together and very similar from a geological perspective. Additionally, with such a small dataset it is often fairly easy to manually clean up all the missing or suspected erroneous input values before these are fed into the machine learning algorithm. In contrast, Rock Solid Images (RSI) have a database of over 2000 wells that have been fully conditioned for geophysical analysis and are available for use in training our models.

The wells that we have access to span many, geologically different, regions from the mid-Norway North Sea to the Barents. They also contain such a volume of real world data (many millions of rows) collected from borehole drilling that it simply is not possible to recreate missing or erroneous values. If you add to this the fact that these interpretations have been conducted by many different people, in some cases going back over 20 years, the data itself is in a challenging state to use as a basis for supervised learning. In this paper we describe the potential benefit that supervised machine learning can bring to streamlining the petrophysical workflow. The contributions of this paper are:

\begin{itemize}
\item An investigation into the applicability of machine learning for the full petrophysical workflow which involves a number of interlinked steps. Previous efforts around machine learning to well log analysis concentrate on one specific step, or portions of a well (see Section \ref{sec:related}) and in this paper we focus on the much more wide ranging workflow across the entire well, feeding our predictions from one stage to the next.
\item Well log data is proprietary and obtaining access to high quality data can be challenging. Previous work has concentrated on a small number of wells that the researchers had access to, with the quality of this data uncertain. In contrast, RSI are world renowned for the quality of their well log data and interpretation. As-such we have access to a very large, high quality, well database to train and test our models with. This is the first time that machine learning has been applied to such high quality well log data in the oil and gas industry.
\item The exploration of a comparison and combination of different machine learning techniques to best optimise our predictions
\item Insight into some of the limitations of common machine learning tools when it comes to HPC machines, such as Crays. We describe work done to mitigate and improve the suitability of some of these tools, to enable them to take full advantage of modern supercomputers.
\item A case study of using HPC for machine learning, as success stories like this are very important to convince the community of the benefits that fusing HPC and ML can deliver.
\end{itemize}

The layout of this paper is as follows, in Section \ref{sec:bg} we highlight some of the related work and state of the art using machine learning for sub-surface data in the oil and gas industry as well as provide more context around the petrophysical workflow that our machine learning algorithms are targeting. In Section \ref{sec:cleaning} we describe the general machine learning approach adopted, some of the technologies used and the initial data cleaning steps undertaken to prepare the data. Sections \ref{sec:minerology}, \ref{sec:porosity}, \ref{sec:saturation} and \ref{sec:lithoclass} describe our use of machine learning to predict the mineralogy composition, porosity, fluid saturation and lithology stages of the petrophysical workflow, before we discuss the challenge of hyper-parameter optimisation and parallelisation work done to enable full use of the Cray XC30 in Section \ref{sec:hyperparams}. Finally, conclusions are drawn and further work discussed in Section \ref{sec:conclusions}. 


\section{Background}
\label{sec:bg}

\subsection{Related work}
\label{sec:related}
Whilst the use of machine learning in the oil and gas industry is still in its infancy, there have been a number of efforts and success stories. An early use of machine learning was in \cite{lithology1992}, where the authors used a neural network, with a single hidden layer, to predict the lithology of wells. Lithology is the general physical characteristics of a rock and in this work the authors aimed to solve a classification problem which predicted whether specific point of the well was one of seven types of rock including limestone, dolomite, sandstone, and shale. They trained their network using density, gamma ray return and neutron porosity input curves from the raw well log data. Whilst they only had very limited amounts of data to train the model with, typically a single well or less, they were still able to demonstrate predictions that picked up the major patterns in lithology. This work concluded that certainly  machine learning, and neural networks in their experience, have a role to play in conditioning well log data, but noted that the human was still critical for quality control and more general interpretation due to limits in their predictions.

The prediction of Clay and Total Organic Content (TOC), which are part of the mineralogy composition, was investigated in \cite{newcastle_paper} where the authors used a combination of well log data with mudstone logs, the latter being the physical samples extracted from the borehole, as input to a neural network. Again, with only one hidden layer, this network used the raw well log input curves gamma ray, resistivity, sonic curve (s-wave), density and borehole tool size as inputs as well as a number of derived curves and the geographical location of the well. In initial experimentation they tested their trained models on a subset of the same wells that their models had already seen, and 85\% of their Clay predictions were within plus or minus 10\% of the truth value, with  95\% of their TOC predictions within plus or minus 1\% of the truth value. In reality this testing, based on data the model had already seen, is of limited use and of more relevance to our work here is that they then ran a series of blind experiments. In this set-up the training and test data was separated and predictions were preformed on test data that the model had never seen before. Whilst their accuracy exploration of these blind experiments was far more limited in detail, from their discussions it was clear that their results matched the truth values fairly closely, and the major patterns of Clay and TOC were picked up by their neural network. 

There are a number of interesting points to highlight in the work done in \cite{newcastle_paper}. Firstly, they included the mudstone logs and from discussions with petrophysicists this adds a significant extra level of complexity. These logs are PDFs of handwritten notes and photographs made during drilling and later analysis, and as such, extracting useful information is far more complex and error prone than the digital well log data. Whilst a petrophysicist does refer to these during their manual interpretation, the use of this data is far more subjective and driven by intuition than the well log data. It is our hypothesis that much of this intuition can be captured in the labelled data that we use to train our models with, and as such the well log data is enough. Whilst the authors of \cite{newcastle_paper} do highlight that the use of a number of wells is advantageous, the total amount of data that they are using to train their models with is still relatively small. This is important because they are using a neural network and that machine learning approach requires all input data to be present. In the paper they do not state whether they rely on raw data without any missing values, manually clean their relatively small amount of data, or only include levels in the well where all values are present, but in our work we do not have the luxury of any of these approaches due to the use of very large amounts of real world data and so face additional challenges.

On the topic of handling real-world, missing data, the Society of Exploration Geophysicists (SEG) held a competition \cite{seg-contest} in 2016 to predict the lithology of rocks, based upon labelled real-world well log data. Solving a classification problem with eight possible inputs curves from the well-logs (depth, gamma ray return, resistivity, photoelectric effect, neutron porosity, density, marine indicator and geographical position), the participants developed models that would predict lithology as accurately as possible. One of the major challenges of this competition was that there were significant amounts of missing data, which is typical for well logs, and the top three winners used a boosted trees approach. Further analysis of the competition results \cite{lithologyxgboost} highlighted that boosted trees were so advantageous here because they are able to handle missing data and still generate predictions, which deep neural networks (DNNs) are unable to do. As such those who relied on DNNs had to perform extra data interpolation, to \emph{fill in the blanks}, and due to the sporadic, and unpredictable, nature of geology this added significant amounts of noise which induced further errors. This is a very important observation for our work, because we have similar challenges when it comes to missing data. In many ways our data is more complex because we have much more of it, and it is interesting that boosted trees performed universally better than the DNNs in \cite{seg-contest}. 

RSI, the industrial collaborators in this research, studied predicting electrical anisotropy in the Barents Sea \cite{electrical_paper} using machine learning. Whilst outside of the core petrophysical workflow, this is useful as it provides key information that can be used to understand regional variations in rock physics properties. They used a multivariate statistics approach to understand which of the raw data log measurements best characterise vertical resistivity measured at the borehole and from this were then able to predict vertical resistivity through a regression model using the Scikit-learn (Sklearn) library \cite{sklearn}. The authors found that this simple approach works reasonable well for predicting vertical resistivity and electrical anisotropy which, whilst the results do not match exactly, are broadly consistent throughout the well. 

As described in this section, the state of the art is that researchers have been focused on small, fairly simple, stages in the petrophysical workflow, such as lithology classification. In this work we are much more ambitious and focus on the entire workflow including aspects such as porosity and fluid saturation which have never before been tackled for the entire well in this context using machine learning. Still, the lessons learnt and approaches described here are very useful to consider and build upon. One important observation from all these papers is that, whilst some authors do use numerical metrics to explore the accuracy of their predictions, without exception all authors mainly use vertical plots of their test wells, with their predicted value curves plotted against the truth values descending by depth. The reason for this is that a single accuracy number can only provide so much information and in a large, deep well, a specific geological entity might have been missed that is extremely important to the petrophysicists, but a single numeric value might not communicate this effectively.

\subsection{Petrophysical interpretation}
\label{sec:petro}
To construct a rock physics model, which is required to accurately understand the geological properties of a well and support activities including oil and gas exploration, a petrophysical interpretation must be performed. This is performed by an experienced petrophysicist and follows a workflow with a number of steps running consecutively, each using the results of previous steps. On average it takes over seven days of human effort to complete the interpretation for one well, and this manually intensive process means that, whilst raw data for many wells is available, the staff effort to process these is often overwhelming. In fact, if raw data for over 20,000 wells is available, it could take in excess of 200 years of effort to provide a complete petrophysical interpretation of each.

\begin{figure}
	\begin{center}
		\includegraphics[scale=0.70]{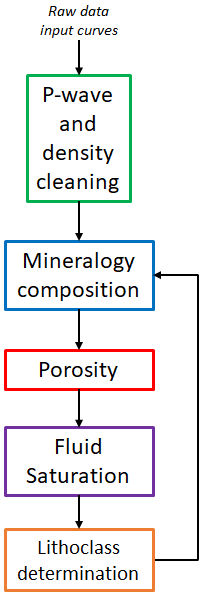}
	\end{center}
	\caption{Illustration of petrophysical interpretation workflow}
	\label{fig-petro-workflow}
\end{figure}

The petrophysical interpretation workflow that we concentrate on in this work is illustrated in Figure \ref{fig-petro-workflow}. Starting from the top, as an initial step the petrophysicist must clean the p-wave and density curves where possible. These two curves are actually a bit of an anomaly because, unlike other input curves, it is often possible for an experienced individual to fairly accurately \emph{fill in the blanks}, although this is time consuming and not always done for the entire well due to time pressure or uncertainty. Once initial cleaning is completed, the petrophysicist then calculates the mineralogy composition of the rock which is what fraction of the rock is one of a number of different minerals. This is then followed by calculating the porosity of the rock, a measure of the empty spaces in the material, and then calculation of the fluid saturation of the rock, determining whether specific reservoirs contain oil, gas or water and what proportion of each. The last stage of the workflow, lithoclass determination, classifies the rock as one of a number of general categories which describe the physical make up of the rock. In each of these stages, the petrophysicist relies on information deduced in previous stages and also often iterates round to further improve their interpretation. Whilst significant skill and experience is required to perform this work, fundamentally the human is performing a pattern matching exercise, albeit a very advanced one. Therefore a key question of the work described in this paper is whether mathematical models can learn to perform these workflow steps and capture the knowledge and experience of the human expert. 

In this work we concentrate on supervised learning, where models are taught based upon labelled training data. The inputs to our mathematical models are raw data input curves coming from the well logs and results of previous workflow steps. Labelled result data trains the model and this comes from previous manual interpretation of the well by petrophysicists from RSI. The well log data we have to work with are text files, formatted in the industry standard, LAS, file format. With one file per well, each row in a file represents measurements at a specific depth in that well, starting from the seabed and then descending down through the rock. Typically, measurements are made every few centimetres, although this does vary and there is an explicit depth measurement that we can rely on. Whilst there are over twenty possible input curves, and many more which can be derived from these raw values, the petrophysicist concentrate on six main features which are depth, gamma ray return, neutron porosity, log of electric deep resistivity, p-wave, and pressure. From experimentation at each stage in the workflow we have found these curves to be sufficient in generating optimal predictions and so concentrate on using these as input curves in our models and in the results presented in this paper.

RSI have a wealth of well log data, and whilst there are over 2000 wells in their database to choose from, in this work we have focused on using wells from the Norwegian and North Sea. The reason for limiting ourselves to a specific region is that different areas contain very different geology and behaviours. For instance the pattern that one would expect to see in the North Sea will different drastically from that seen in the Barrents. As such, concentrating on a specific area means that, whilst geological formations do still vary within a region, wells will behave much more similarly than they would across multiple regions. Due to the significant amount of high quality data, we have been able to pick a region which is of huge commercial value to RSI, the data is abundant and high quality, and importantly the region is one which the petrophysicist involved in this project have experience in. This last point is very important because the truth value that we train our models with are themselves a manual interpretation. Therefore discrepancies between truth and prediction might not necessarily mean that our models are wrong and so having the expertise to interpret and contextualise our results is crucial. Focusing on RSI's wells from the Norwegian and North Sea, we use over one hundred wells in this work which provides just over 1.2 million rows of data that we can use to train and test our models. In terms of the amount of data, this is by far the largest data set used so far in machine learning for well log conditioning.

Raw well log data is captured by instruments run down boreholes and this real world data is challenging, both in terms of missing values and also potential noise. Out of these two concerns, it is the missing data that is most problematic. This is because it is very common for a drill not to record values for a variety of results, for instance due to the expense of gathering data for the entirety of the well, a well casing point or reliability issues with the tool. Furthermore, the input curves vary significantly throughout a well, which is the nature of geology and as such petrophysicists try to avoid simple interpolation to fill in the missing values as this often adds significant uncertainty and error, which corresponds to observations made in \cite{lithologyxgboost}. As such petrophysicists tend to live with missing values and perform their interpretation in the presence of these.

When it comes to machine learning, as highlighted in \cite{lithologyxgboost}, missing data is a challenge to some models. Neural networks, arguably one of the most popular machine learning approaches, require all the input data to be present in order to make a prediction. To this end, building on the experiences of \cite{lithologyxgboost}, in this work we use boosted trees \cite{boosted-trees}. Otherwise known as gradient boosting, this approach relies on the idea of decision tree ensembles where a model consists of a set of classification or regression trees and features of the problem are split up amongst tree leaves. Each leaf holds a score associated with that feature and as one walks the tree, scores are combined which then form the basis of an overall prediction answer. Generally speaking, usually a single tree is not sufficient for the level of accuracy required in practice, and so an ensemble of trees, where the model sums the prediction of multiple trees together, is used. As one trains a boosted trees model, the trees are built one at a time, with each new tree helping to correct the errors made by previously built trees. This is one of the factors that makes boosted trees so powerful and they have been used to solve many different machine learning challenges \cite{boosted1}\cite{boosted2}\cite{boosted3}. Most importantly in this work, boosted trees handle missing data values, as the corresponding tree is simply down weighted, but a major challenge is that they are more difficult to tune and highly sensitive to their hyper-parameters \cite{boosted-sensitive}


\section{General approach and initial data cleaning}
\label{sec:cleaning}

For this work we use Python 3 and in our machine learning scripts we initially load up our well log data files into a Pandas data frame which makes it trivial to perform data manipulation in code. Throughout the experiments detailed in this paper the data is randomly split up on a well by well basis, with 80\% of the wells in the training set and 20\% of the wells in the test set. This means that, without exception, our experiments are blind and the trained models are tested, sight unseen, by predicting on the test set and then comparing against the true values in this test data to determine prediction performance (accuracy). We use the XGBoost library \cite{xgboost} which is an open source software framework aiming to provide a scalable, portable and distributed gradient boosting library for Python and numerous other languages. Whilst this is mature when run on CPUs, one of the challenges we initially faced was that XGBoost is rather buggy when it comes to running on GPUs. This is likely due to the large amount of raw data that we have available for training, because after one or two runs of our model the GPU runs out of memory and raises an error. From exploration we found that there is a documented issue around memory leakage with the library on GPUs and at the time of writing this is currently outstanding.

Hence, for this work, we limited our models to running on CPU and for this we used ARCHER, a homogeneous Cray XC30 and used the Anaconda module as a base Python setup. The XGBoost library has been parallelised with OpenMP, which is far more mature than their GPU implementation and we run a single boosted trees model per NUMA region (12 cores of Ivy Bridge in the XC30.) On average it takes around ten to fifteen minutes to train each single model, however training times grow very significantly when we increase the number of input curves. The ability to thread over 12 cores is useful here, otherwise the runtime would have been far longer. Additionally, the large amount of memory provided by the XC30 meant we could fit all the raw data into RAM. Once trained, the model takes around a second to make predictions on each test well. In terms of our aim, \emph{reducing petrophysical interpretation time down from over 7 days to 7 minutes}, it is this prediction, or inference, time rather than the training time that counts, because the assumption is that the models used for interpreting a well have already been trained and validated across RSI's regional data-set.

In all of our boosted trees model runs, we use the Root Mean Square Error (RMSE) as the evaluation metric for XGBoost. This, along with the choice of boosting algorithm is provided as configuration options to the XGBoost API and we found that the default, \emph{gbtree}, boosting algorithm works best. There are seven hyper-parameters that control the training of the model and, due to the sensitive nature of boosted trees, setting these is not trivial. Not only do the most appropriate hyper-parameter values vary on a model by model basis, but also whenever we experiment with using the same model in different ways, such as changing the input curves, new hyper-parameters need to be found. See Section \ref{sec:hyperparams} for a detailed discussion of how we pick the appropriate settings in this work. 

The first step in Figure \ref{fig-petro-workflow}, and a preliminary stage in the petrophysical workflow, is the cleaning of p-wave and density curves. The petrophysicists are actually doing two activities here, firstly they are cleaning the curves to remove any obvious errors and secondly attempting to fill in any blanks in these curves. Due to the challenging nature of real-world geology, this is often a time consuming process and requires significant expertise. We investigated whether machine learning could be used to accurately clean the two curves and as an input to our boosted trees models, we use depth, gamma ray return, neutron porosity, the log of the deep resistivity, and the original density and p-wave curves.

\begin{figure}
		\includegraphics[scale=0.50]{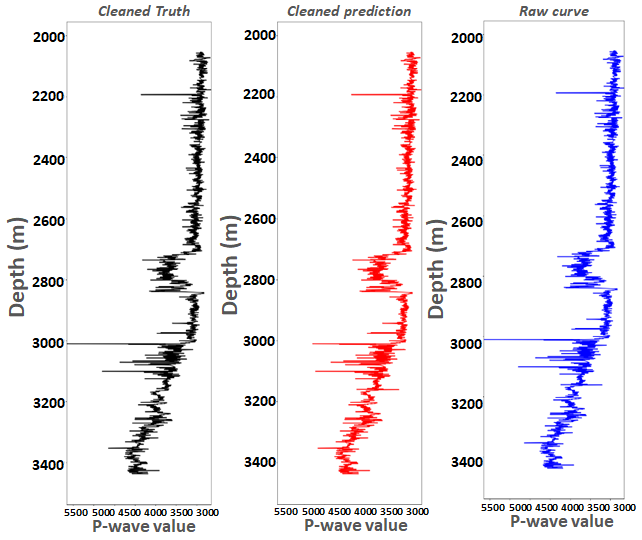}
	\caption{Cleaning of the p-wave curve, the manually cleaned curve is the left plot (black curve), our model prediction is the middle plot (red curve) and the original raw data curve is the plot on the right (blue curve)}
	\label{fig-cleaning-vp}
\end{figure}

Figure \ref{fig-cleaning-vp} illustrates the results of our model cleaning the p-wave curve on one of our test wells. Plots such as the three in Figure \ref{fig-cleaning-vp} are a standard way of presenting the values of curve(s) in a well. The vertical axis is depth, with the top of the plot representing the seabed floor, and depth increases as we go down the plot and effectively descend through the rock. The horizontal axis is the range of values of the curve(s) being presented. Some wells require far more work than others and, whilst the predicted cleaned curve which comes from our model (red) in the middle plot of Figure \ref{fig-cleaning-vp} looks to closely match the manually cleaned curve (black) on the left (the truth), in-fact it can be seen that there is little to do here because the original curve (blue, right plot) is complete and the petrophysicist made very few modifications. However this is still useful to note for two reasons, firstly the reader can see the reduction in spike about two thirds the way down the well, at 3000m which our model performed but the manual interpreter did not. Upon further investigation the petrophysicist deduced that actually our model is most accurate here and the manual interpretation should have included a similar adjustment. Even though this well is fairly simple for our model to work with, it is also important to highlight that our approach has not applied false adjustments which would make the cleaned curve worse. Furthermore, the observation we made in Section \ref{sec:related} applies here too, where the best way of comparing prediction accuracy is by plotting by depth and comparing them. Not only would a single metric number likely hide the reduced spike at 3000m, but also bearing in mind the accuracy of the original curve this would mean little in terms of the accuracy of the prediction.

\begin{figure}
	\begin{center}
		\includegraphics[scale=0.50]{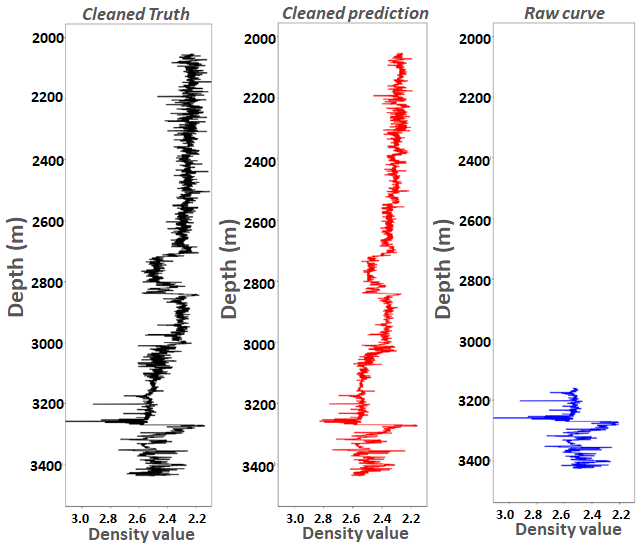}
	\end{center}
	\caption{Cleaning of the density curve, the manually cleaned curve is the left plot (black curve), our model prediction is the middle plot (red curve) and the original raw data curve is the plot on the right (blue curve)}
	\label{fig-cleaning-rhob}
\end{figure}

Where things become far more interesting is in Figure \ref{fig-cleaning-rhob}, which illustrates the cleaning of the density curve for the same well. In this case things are far more challenging because the vast majority of the raw curve is missing (right plot, blue curve.) It is a much more time intensive process for the petrophysicist to reconstruct the entire curve based on the small section present at the bottom of the well, and this is where the use of machine learning can be of most benefit. It can be seen from comparing the manually interpreted truth (left plot, black curve) against our model's prediction (middle plot, red curve) in Figure \ref{fig-cleaning-rhob} that our model was able to predict the density curve for the rest of the well matching fairly closely against the manually interpreted version. Whilst this is not quite a perfect match, it falls within the general bounds provided by the petrophysicists, as their own interpretation contains some degree of error, and is considered sufficient for use in later stages of the petrophysical workflow. Bearing in mind that cleaning such curves is challenging and time-consuming for a manual interpreter, the fact that our model can generate these fairly accurately in a matter of seconds is very useful.

\section{Mineralogy}
\label{sec:minerology}
Once the p-wave and density input curves have been cleaned then the petrophysicists can start on the main petrophysical interpretation. The first stage here is to determine the mineralogy composition of rock and effectively they are deducing what fraction of the rock is one of thirteen minerals. For each row of data, corresponding to a level in the well and most often on a metre by metre basis, the overall values of minerals at that level will sum up to a total of one. For instance a level might have 0.35 Clay, 0.2 Calcite, 0.15 Coal and 0.3 Quartz, which sums to 1.0. 

\begin{figure}
\begin{center}
\begin{tabular}{ | c c | }
\hline
 Mineral & RMS error \\ \hline
 Clay & 0.136427\\  
 Quartz & 0.145153\\  
 Calcite & 0.049276\\
 Pyrite & 0.004348\\
 Dolomite & 0.011489\\
 Coal & 0.050087\\
 TOC & 0.000394\\
 Anhydrite & 0.003198\\
 Volcanic & 0.005829\\
 Feldspar & 0.023668\\
 Siderite & 0.000772\\
 Halite & 0.000514\\
 \hline
\end{tabular}
\end{center}
\caption{Model prediction accuracy across all wells in the test set}
\label{fig-mineral-accuracy}
\end{figure}

As described in Section \ref{sec:petro}, from previous work, experimentation and domain knowledge, we know that there are six crucial input curves, which the petrophysicists themselves use, when it comes to predicting the mineralogy and petrophysical workflow in general. These are used as the input to our boosted regression trees model, including the cleaned p-wave and density curves of Section \ref{sec:cleaning} and we train separate models for each mineral. Figure \ref{fig-mineral-accuracy} illustrates the prediction accuracy for separate models trained for each mineral, using the Root Mean Square Error (RMSE) for each type of mineral across the entire test set. For brevity, the rest of this section concentrates on only three minerals, Clay, Quartz and Calcite, as the patterns and behaviours exhibited here also apply to the other minerals. 

\begin{figure}
	\begin{center}
		\includegraphics[scale=0.50]{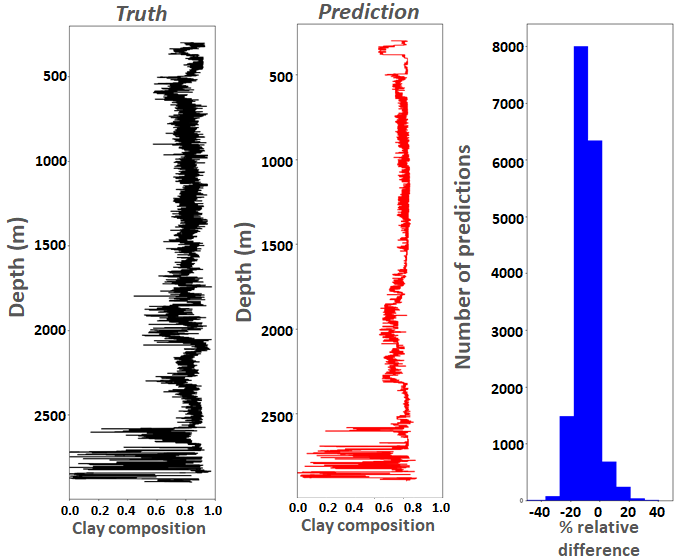}
	\end{center}
	\caption{Fraction of Clay by depth, left plot (black curve) illustrates the manually interpreted truth, our model's prediction is the middle plot (red curve) and the histogram on the right illustrates the number of predictions that fall within a specific percentage accuracy relative to the truth}
	\label{fig-clay-minerology}
\end{figure}

Figure \ref{fig-clay-minerology} illustrates our model's Clay prediction, where the middle plot (red curve) is our prediction and should be compared against the left plot (black curve) which is the manually interpreted, true value. It can be seen that, whilst the prediction picks up the majority of the shape of the Clay curve, these are not an exact match, especially when it comes to the magnitude. It is however important to be aware of two factors: Firstly it takes over eight hours for an experienced petrophysicist to produce the mineralogy curves, whereas our trained model generates them in the order of a few seconds, and secondly the \emph{truth} is itself an interpretation and hence there is some degree of subjectivity. The right most plot of Figure \ref{fig-clay-minerology} illustrates the number of predictions that fall within a specific percentage accuracy range range relative to the truth. It can be seen that the vast majority of our model's predictions fall within plus or minus 20\% of the truth value. When bearing in mind the work done in \cite{newcastle_paper}, as described in Section \ref{sec:related}, their wells tended to be within plus or minus 10\% of the truth value, but crucially for this measure they were testing on wells that their model had already seen, whereas in our approach we are testing on wells that the model has never seen previously and-so it is a much more difficult problem.

\begin{figure}
	\begin{center}
		\includegraphics[scale=0.50]{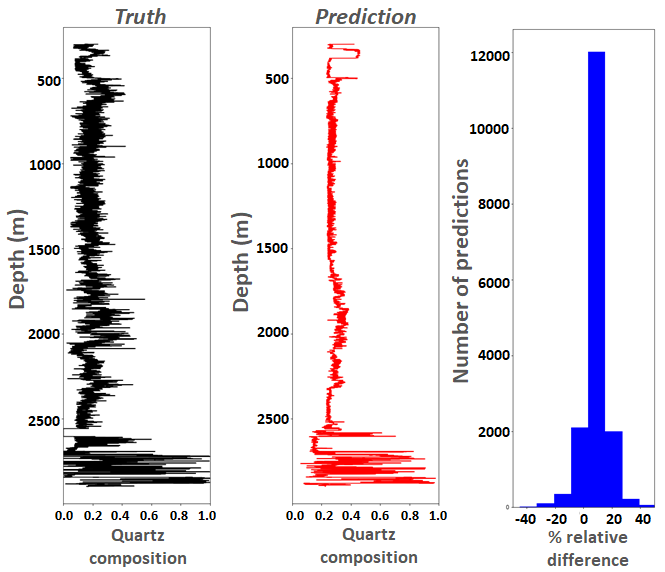}
	\end{center}
	\caption{Fraction of Quartz by depth, left plot (black curve) illustrates the manually interpreted truth, our model's prediction is the middle plot (red curve) and the histogram on the right illustrates the number of predictions that fall within a specific percentage accuracy relative to the truth}
	\label{fig-quartz-minerology}
\end{figure}

\begin{figure}
	\begin{center}
		\includegraphics[scale=0.50]{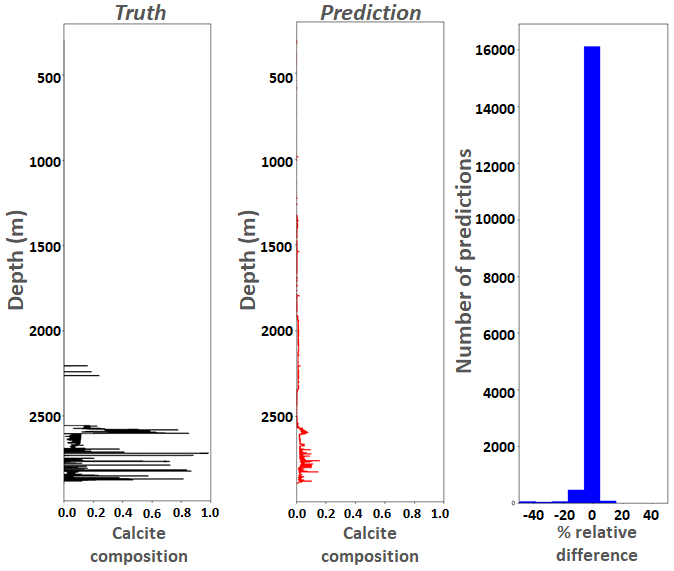}
	\end{center}
	\caption{Fraction of Calcite by depth, left plot (black curve) illustrates the manually interpreted truth, our model's prediction is the middle plot (red curve) and the histogram on the right illustrates the number of predictions that fall within a specific percentage accuracy relative to the truth}
	\label{fig-calcite-minerology}
\end{figure}

Figures \ref{fig-quartz-minerology} and \ref{fig-calcite-minerology} illustrate the predictions against true values for Quartz and Calcite respectively. Quartz is a similar story to the Clay prediction, where the general shape is picked up but the magnitude can deviate at specific points. From discussions with the petrophysicists they have identified that our Clay predictions are more accurate than Quartz predictions. The Calcite predictions are interesting as, from Figure \ref{fig-mineral-accuracy} and the differences plot in the right of Figure \ref{fig-calcite-minerology}, one might assume that this would be by far the most accurate prediction out of the three minerals we are focusing on in this paper. However it goes back to the point made in Section \ref{sec:related}, that these numbers are heavily influenced by the fact that the majority of the well has zero Calcite, which our model picks up. When Calcite starts to appear towards the bottom of the well then our model struggles to predict it accurately. Once again, our model is identifying that there is some Calcite present but struggles with the magnitude and this illustrates the use of studying these depth plots, because patterns and inconsistencies can be highlighted that a single error numeric value masks. 

\begin{figure}
	\begin{center}
		\includegraphics[scale=0.50]{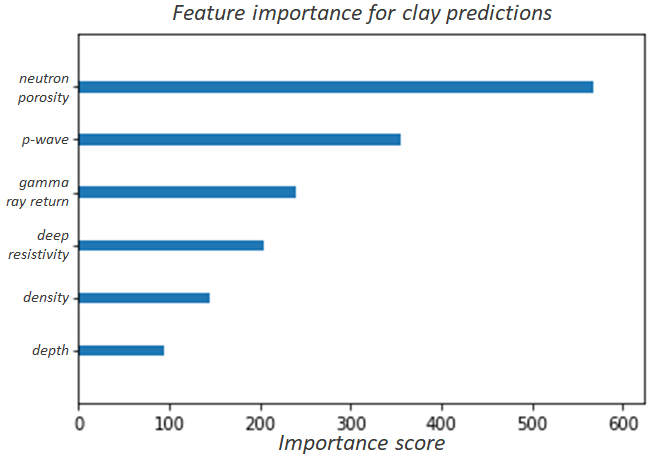}
	\end{center}
	\caption{Weight feature importance scores for Clay prediction, as reported by the XGBoost library}
	\label{fig-clay-feature-importance}
\end{figure}

From detailed investigation we found that the quality of mineral prediction depends heavily on the available data. But crucially not all input data is of equal importance. Figure \ref{fig-clay-feature-importance} illustrates the weight feature importance score for each input curve for Clay predictions, as reported by the XGBoost library. It is a very similar story for the prediction of other minerals, and it can be seen that by far the most important feature is the neutron porosity input curve, followed by the pressure and gamma ray. This is important information, because in the well we have studied in this paper, between 1500m and 2300m, the neutron porosity curve is entirely missing and the availability of the gamma ray curve is sporadic. Hence, whilst the boosted trees model is able to still generate a prediction regardless of this missing data, the first and third most important features are missing in this range. From looking at the Clay and Quartz predictions of Figures \ref{fig-clay-minerology} and \ref{fig-quartz-minerology}, it can be seen that the prediction is especially inaccurate in this range and the missing input data, whilst the model can still make a prediction, is limiting the accuracy.

A key question is why our Calcite model struggled with the curve magnitude towards the bottom of the well. The main reason for this is that geology is inherently biased, where some minerals such as Clay and Quartz are simply seen much more regularly than others. Hence these models have more experience in how to handle and deal with the more abundant minerals and can therefore make a better job of predicting them. The fact that the vast majority of wells contain large sections of zero Calcite means that the model biases no Calcite over some being present and hence it has a tendency to under predict or even miss the Calcite all together, especially if there is some degree of uncertainty. 

\subsection{Inclusion of formations}

An interesting observation of Figure \ref{fig-clay-feature-importance} is that depth is the least important feature. This is very interesting because the petrophysicists use depth for providing context to the other curves. However, depth is really just a symptom of the fact that they are actually concerned with the underlying geological formation. This makes a lot of sense, because different geology will result in different input curve values that can mean the same thing. A number of the research activities described in Section \ref{sec:related} use geographical location as an input curve, but we found that this gave no improvement to the overall prediction. However, effectively what we were trying to do was include a notion of the underlying geological formations and these are not directly linked to the geographical location. This is because in one area the geology can change significantly whereas in other areas formations can remain very stable. Certainly the wells we are using from across the Norwegian and North Sea region do encounter many areas of changing formations and so an important experiment was to investigate whether including these formations in our data improved the quality of prediction or not. Whilst the formations themselves are not included in the raw data files, these are freely available from Norwegian Petroleum Directorate's website (NPD) \cite{npd}. Amongst other things, the website contains a database of well information which covers the Norwegian sea and based on this there are twenty seven possible formations. Per row in the well logs, formations are mutually exclusive, so there is exactly one formation per level. We represent each formation as an extra numeric input curve, one being that formation is present and zero the formation is absent.

\begin{figure}
\begin{center}
\begin{tabular}{ | c c c | }
\hline
 Mineral & \thead{No formations \\ RMS error} & \thead{Formations \\ RMS error} \\ \hline
 Clay & 0.136427 & 0.132107\\  
 Quartz & 0.145153 & 0.140282\\  
 Calcite & 0.049276 & 0.049098\\
 Pyrite & 0.004348 & 0.004628\\
 Dolomite & 0.011489 & 0.017524\\
 Coal & 0.050087 & 0.040587\\
 Toc & 0.000394 & 0.000394\\
 Anhydrite & 0.003198 & 0.003360\\
 Volcanic & 0.005829 & 0.005829\\
 Feldspar & 0.023668 & 0.024003\\
 Siderite & 0.000772 & 0.001289\\
 Halite & 0.000514 & 0.001185\\
 \hline
\end{tabular}
\end{center}
\caption{Model prediction accuracy across all wells in the test set with and without geological formation information}
\label{fig-mineral-accuracy}
\end{figure}

\begin{figure}
	\begin{center}
		\includegraphics[scale=0.50]{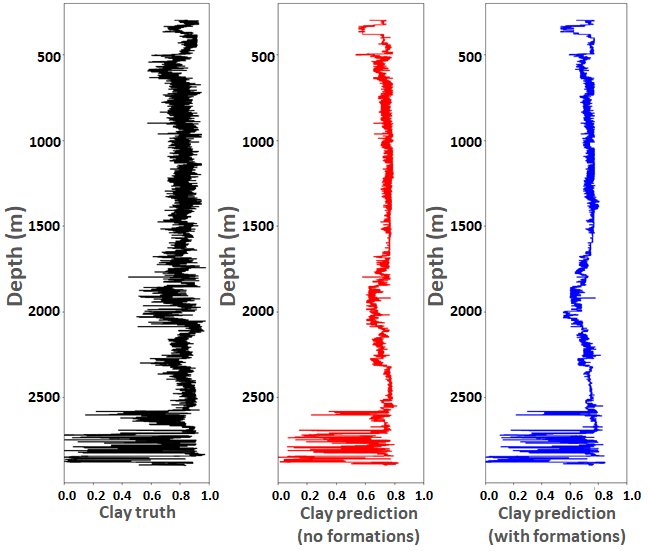}
	\end{center}
	\caption{Fraction of Clay by depth, left plot (black curve) illustrates the manually interpreted truth, our model's prediction without formations is the middle plot (red curve), and our model's prediction with formations is the right plot (blue curve)}
	\label{fig-clay-formations-plot}
\end{figure}

Figure \ref{fig-mineral-accuracy} illustrates the RMS error across our test set for each mineral's model when the models are trained and tested with and without formation information. It can be seen that the inclusion of formations makes little difference: In some situations it slightly improves the overall accuracy and in other situation the prediction is slightly worse. The same conclusions can be drawn from Figure \ref{fig-clay-formations-plot}, which illustrates the Clay prediction. The manually interpreted (truth) value is the left plot, our previous model prediction without formations is in the middle (red curve) and the prediction of our new model which includes formations is on the right (blue curve). From this plot it can be seen that, whilst there are some minor differences, the inclusion of formations has very little overall impact in a systematic manner and the same is true for all minerals. This was a very interesting result because the petrophysicists thought that formations could make a significant improvement to the overall mineralogy prediction, whereas in reality when examining the predictions they found very little qualitative change. This observation was further strengthened by an exploration of the boosted trees feature importance report, where the ranking of the raw input curves of Figure \ref{fig-clay-feature-importance} remain unchanged and formations are considered less important by the model than the well log raw data input curves.

\subsection{Alternative machine learning approaches}
Another important question regarding improving the accuracy of our mineralogy predictions is the machine learning method to use. We chose boosted trees due to the significant amount of missing data, and whilst it is possible to perform some interpretation on the p-wave and density curves to fill in the blanks, this is not possible for the other input curves. But, for experimentation purposes, if we limit ourselves to the rows of the wells where all input data is present, this also opens up the possibility of using neural networks to do our prediction. Generally speaking, the restriction of processing only parts of the well where all input curves are present means that this is not particularly useful to the petrophysicists in the real-world, because on average only around half of each well can be predicted. This is still useful to explore and understand because a question is, if we were in an ideal world with complete data, then would other methods provide improved prediction capabilities? 

Figure \ref{fig-clay-nn-accuracy} illustrates the accuracy of Clay prediction based on models built using a number of different machine learning methods, processing just on levels in the wells that contain all the input data. The \emph{base} entry represents the RMS error of the Clay prediction across our test wells as described previously. The first method we tried was a Multi Layer Perceptron (MLP) using the Sklearn toolkit for regression. From experimentation we found that two hidden levels, each with 20 neurons, and using a \emph{relu} activation layer along with \emph{adam} solver, an activation hyper-parameter of 0.1, and 1000 iterations gave the best prediction performance. The Deep Neural Network (DNN) entry of Figure \ref{fig-clay-nn-accuracy} represents a deep neural network using the PyTorch machine learning framework. In contrast to the MLP model, PyTorch provides far more control over the general configuration and each of the layers. For this DNN we used four hidden layers, each with thirty neurons, 500 epochs, a batch size of 2000 and learning rate of 0.001. We are using a softmax activation layer and the mean squared error loss function. The last two entries in Figure \ref{fig-clay-nn-accuracy} refer to our existing boosted trees model, the first of these is only trained and tested on levels in the wells with all the data. The second boosted trees entry, \emph{Boosted trees (missing data for training)}, is trained on all levels in the wells of the training set, regardless of whether they contain missing data, but only predicts on levels in the test set wells where all input curves are present. These two configurations link back to the observations made about mineralogy predictions earlier, where the accuracy of prediction is impacted by missing input data, especially if these are important features. In theory predicting only on levels which contain all the input curves will provide more accuracy and the question was whether it is beneficial to still train the model on all the data in the training set, even if this contains missing data, as the model will experience a wide variety of data. The results of Figure \ref{fig-clay-nn-accuracy} illustrate that it is beneficial to use as much data as possible when training the boosted trees model.


\begin{figure}
\begin{center}
\begin{tabular}{ | c c | }
\hline
 Method & Clay prediction RMS error\\ \hline
 Base & 0.136427 \\
 MLP (Sklearn) & 0.1772 \\  
 DNN (PyTorch) & 0.0651 \\  
 Boosted trees & 0.1033 \\
 \makecell{Boosted trees \\ (missing data for training)} & \makecell{0.0838\\} \\
 \hline
\end{tabular}
\end{center}
\caption{Model prediction accuracy using different methods across test wells for Clay prediction}
\label{fig-clay-nn-accuracy}
\end{figure}

It can be seen in Figure \ref{fig-clay-nn-accuracy} that, when we limit our Clay predictions to levels in the well where all input curves are present, the DNN is by far the most accurate approach. Interestingly the MLP is the least accurate, but it is the ability to tune the configuration of the neural network model here that makes a big difference. It can also be seen that boosted trees predictions are better than the base prediction when we exclude levels in the test well that have missing input values. This is to be expected, but it is interesting that the prediction improves when we include the partial data in the training set. Note that due to the significant amount of missing data, a depth plot of the prediction vs truth curves is not particularly useful in this situation because so many points are missing and hence key features are lost.

\section{Porosity}
\label{sec:porosity}
Porosity of the rock measures the void, or empty, spaces that are present. In our context this is reported as a fraction between 0 and 1 of the volume of voids over the total volume. We initially trained a boosted trees regression model on the measurements directly from the raw data, as they come from the borehole. The idea was to have a base model that doesn't require any cleaning of p-ware or density curves, or previous petrophysical stages, as this helps us to understand whether this stage in the workflow can be performed separately of whether it requires data cleaning and/or mineralogy composition. Bearing in mind some of the challenges around accurate mineralogy prediction, this also enabled us to understand how useful our mineralogy predictions are when it comes to using them as inputs to further stages in the workflow.

Figure \ref{fig-porosity-accuracy} illustrates the RMS error for our boosted trees regression model across the entire test set, run on the raw curves (i.e. no p-wave or density cleaning) with and without mineralogy information provided as additional inputs to the model. When it came to providing the mineralogy we explored a number of different options ranging from supplying just Clay, the most abundant mineral, to supplying a subset of the minerals, to providing the full mineralogy. It can be seen that, whilst providing mineralogy improves the accuracy, the difference between the clay only and full mineralogy cases is small. If anything, providing Clay, Quartz and Calcite rather than just Clay or the full mineralogy is slightly beneficial. This is most likely because these are the three most common minerals, so not only do these have the greatest impact generally on the porosity, but also our mineralogy regression model as described in Section \ref{sec:minerology} is most confident predicting these minerals.

\begin{figure}
\begin{center}
\begin{tabular}{ | c c | }
\hline
 Configuration & RMS error \\ \hline
 No mineralogy & 0.045896 \\  
 Clay only & 0.040109 \\ 
 Clay, Quartz, Calcite & 0.039610 \\  
 Full mineralogy & 0.040583 \\
 \hline
\end{tabular}
\end{center}
\caption{Porosity prediction error rate and presence of mineralogy}
\label{fig-porosity-accuracy}
\end{figure}

Figure \ref{fig-porosity-accuracy-cleaned-curves} illustrates the RMS error against use of mineralogy when the cleaned p-wave and density curves are used instead of their raw counterparts. It can be seen that this significantly improves the accuracy of all configurations, and the inclusion of our previous mineralogy predictions is still advantageous in reducing the prediction error. In contrast to using the raw curves, when using the processed curves in-fact including the full mineralogy, predictions for all thirteen minerals, is the optimal configuration to use, although the difference in error between that and using only Clay, Quartz and Calcite is fairly small. 

\begin{figure}
\begin{center}
\begin{tabular}{ | c c | }
\hline
 Configuration & RMS error \\ \hline
 No mineralogy & 0.0316558 \\ 
 Clay only & 0.026754 \\
 Clay, Quartz, Calcite & 0.025536 \\  
 Full mineralogy & 0.023891 \\
 \hline
\end{tabular}
\end{center}
\caption{Porosity prediction error rate and presence of mineralogy for cleaned p-wave and density curves}
\label{fig-porosity-accuracy-cleaned-curves}
\end{figure}

\begin{figure}
	\begin{center}
		\includegraphics[scale=0.50]{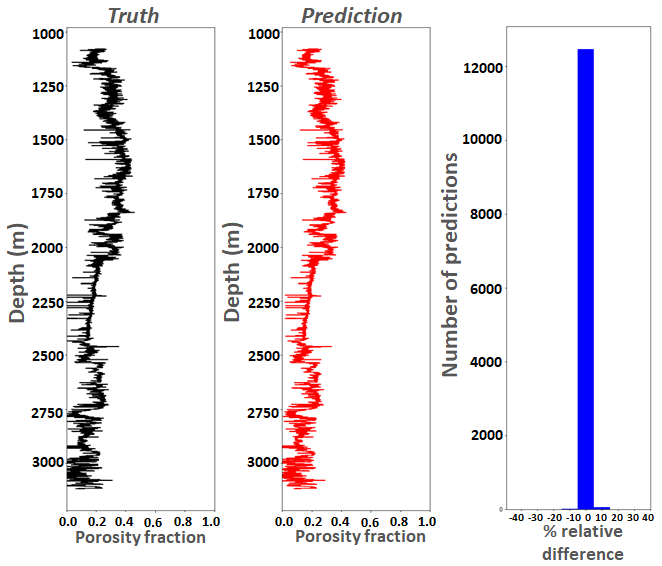}
	\end{center}
	\caption{Fraction of porosity by depth using cleaned p-wave and density curves and full mineralogy information. The left plot (black curve) illustrates the manually interpreted truth, our model's prediction is the middle plot (red curve) and the histogram on the right illustrates the number of predictions that fall within a specific percentage accuracy relative to the truth}
	\label{fig-porosity-full-minerology}
\end{figure}

\begin{figure}
	\begin{center}
		\includegraphics[scale=0.50]{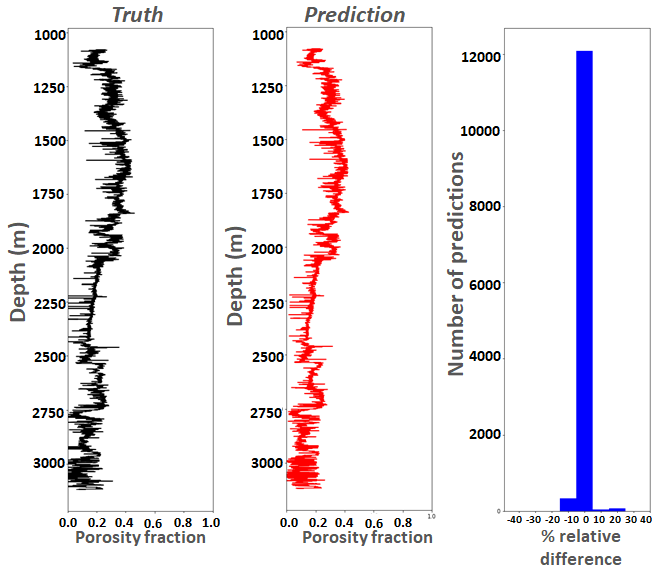}
	\end{center}
	\caption{Fraction of porosity by depth using cleaned p-wave and density curves but no mineralogy information. The left plot (black curve) illustrates the manually interpreted truth, our model's prediction is the middle plot (red curve) and the histogram on the right illustrates the number of predictions that fall within a specific percentage accuracy relative to the truth }
	\label{fig-porosity-no-minerology}
\end{figure}

Clearly, using the cleaned p-wave and density curves is beneficial here and Figure \ref{fig-porosity-full-minerology} illustrates the porosity prediction by our model for a single well in the test set. The middle plot (red curve) is our prediction using the processed curves and full mineralogy, against the manually interpreted, truth, value (left plot, black curve). The RMS error for this well's prediction is 0.022392, so fairly average for the wells in the test set. From the histogram on the right of Figure \ref{fig-porosity-full-minerology}, it can be seen that the vast majority of predictions are with in 10\% of the truth value and the petrophysicists consider that the prediction matches very closely here. Figure \ref{fig-porosity-no-minerology} illustrates the same experiment, where no mineralogy information is fed to the model, but still with the processed curves. For comparison the RMS error here is 0.033002, so again a fairly average error for the wells in the test set. Interestingly for this well, removing the mineralogy information has made the prediction RMS error go from slightly better than average to slightly worse than average. From comparing the difference histograms in Figures \ref{fig-porosity-full-minerology} and \ref{fig-porosity-no-minerology}, it can be seen that removing mineralogy results in predictions that are less accurate, as would be expected from the errors reported in Figure \ref{fig-porosity-accuracy-cleaned-curves}. Whilst a detailed comparison of the prediction curves does highlight some qualitative differences, these are considered minor by the petrophysicists and also within the acceptable accuracy range that they have dictated. This is an important result because it means that, in-fact, whilst the porosity calculation must use the cleaned p-wave and density curves, the impact of not using the mineralogy prediction is fairly minimal. From this section we can conclude that predicting porosity is highly accurate and reliable, even though our mineralogy predictions contained some errors, they are still accurate enough to be used by this stage if needed.

\section{Fluid Saturation}
\label{sec:saturation}
When calculating the fluid saturation of rock, the petrophysicist is focusing on specific reservoirs that could contain hydrocarbons (oil or gas), which is what they are looking for, or water which is uninteresting to them. Unfortunately for the oil and gas industry, water is much more common in these reservoirs than hydrocarbons, so they need to be able to accurately determine the nature of the fluid. We use the six normal input curves, with cleaned p-wave and density, in combination with the porosity from predictions in Section \ref{sec:porosity} and full mineralogy from predictions in Section \ref{sec:minerology}. We are again using boosted trees regression, with three separate models. One trained to calculate the water saturation of the rock, another to calculate the oil saturation of the rock and the third to calculate the gas saturation of the rock.

\begin{figure}
	\begin{center}
		\includegraphics[scale=0.40]{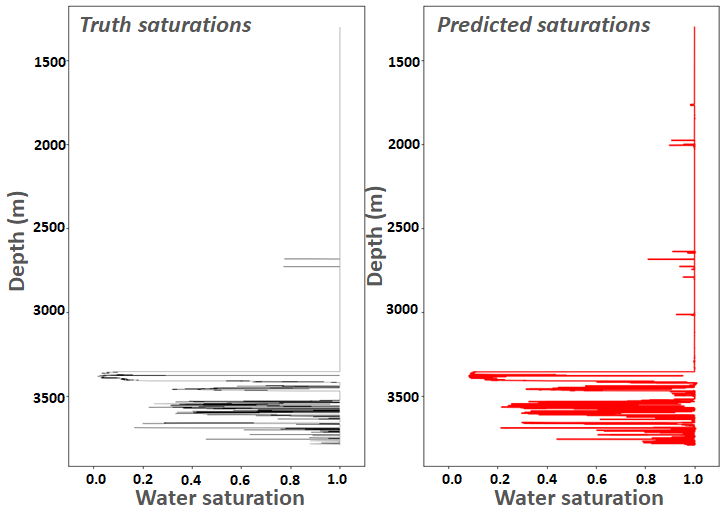}
	\end{center}
	\caption{Water saturation by depth, manually interpreted (truth) saturation in the plot on the left (black curve) and our prediction in the plot on the right (red curve)}
	\label{fig-simple-water-saturation-by-depth}
\end{figure}

\begin{figure}
	\begin{center}
		\includegraphics[scale=0.40]{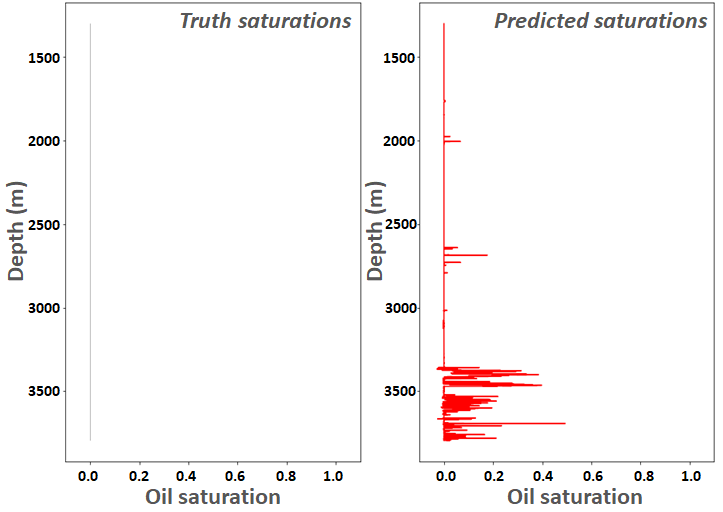}
	\end{center}
	\caption{Oil saturation by depth, manually interpreted (truth) saturation in the plot on the left (black curve) and our prediction in the plot on the right (red curve)}
	\label{fig-simple-oil-saturation-by-depth}
\end{figure}

\begin{figure}
	\begin{center}
		\includegraphics[scale=0.40]{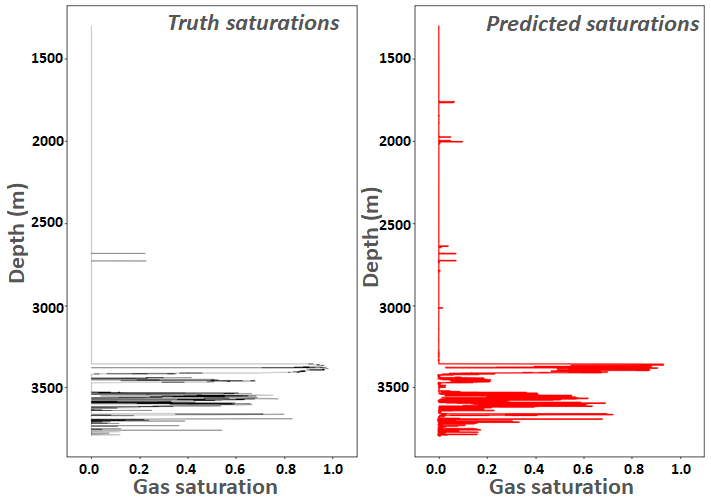}
	\end{center}
	\caption{Gas saturation by depth, manually interpreted (truth) saturation in the plot on the left (black curve) and our prediction in the plot on the right (red curve)}
	\label{fig-simple-gas-saturation-by-depth}
\end{figure}

Figure \ref{fig-simple-water-saturation-by-depth} illustrates the water saturation by depth for both the manually interpreted, truth, saturation on the left (black curve) and our prediction on the right (red curve). From comparing these images it can be seen that our model picks up most of the water, but has a tendency to under predict water at specific points in the well (the plot ranges from 0\% water on the left to 100\% water on the right). Figures \ref{fig-simple-oil-saturation-by-depth} and \ref{fig-simple-gas-saturation-by-depth} illustrate the prediction of oil and gas respectively by depth. The reader can see that there are some major issues with these predictions, for instance whilst it is known that there is no oil in the well, our model predicts oil, and the other model trained on gas has a tendency to under-predict. The reason for this is that the markers for oil and gas are actually very similar, and the petrophysicists are not able themselves to use only well log data to accurately identify whether it is oil or gas.

\begin{figure}
	\begin{center}
		\includegraphics[scale=0.40]{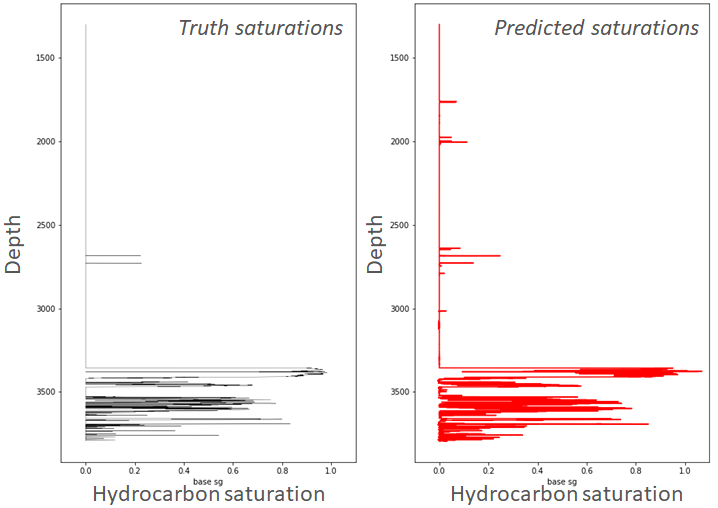}
	\end{center}
	\caption{Hydrocarbons saturation by depth, manually interpreted (truth) saturation in the plot on the left (black curve) and our prediction in the plot on the right (red curve)}
	\label{fig-simple-hc-saturation-by-depth}
\end{figure}

Figure \ref{fig-simple-hc-saturation-by-depth} illustrates the combination of predictions from our oil and gas models in Figures \ref{fig-simple-oil-saturation-by-depth} and \ref{fig-simple-gas-saturation-by-depth}. Comparing the manually interpreted, truth, hydrocarbon values in the plot on the left (black curve) against the predicted hydrocarbon values in the plot on the right (red curve), it can be seen that, whilst this prediction is still not perfect, it is far more accurate than when the predictions were split out. A question we had going into this work was whether the extra intuition required for distinguishing between oil and gas could be captured by the labelled training data. From experimentation we have found that, whilst it is possible to train our models to identify water from hydrocarbons, it is not possible to sub categorise the hydrocarbons group into oil or gas. It is our conclusion that, based on well log data alone, the best one can hope for is water vs hydrocarbons, which itself is extremely useful.

 So far we have used a regression model to predict the fluid saturation, which outputs a number representing the fraction between 0 and 1 of saturation at each depth. A limitation of this is that our predictions are rarely 100\% water saturation or 0\% hydrocarbon saturation, but the situation where the rock is entirely saturated with water is by far the most common configuration in the real world. Instead our water prediction has a tendency to wiggle around the 100\% water saturation point, as can be seen in Figure \ref{fig-simple-water-saturation-by-depth}. Additionally, up until this point we have assumed two separate models, one to predict water saturation and the one to predict hydrocarbon saturation. However, this is not actually necessary because the amount of water plus the amount of hydrocarbons must be equal to 1 at each depth in the well. Effectively this means we can predict only the water saturation, and then derive the hydrocarbon saturation by then inverting it.

We therefore decided to test a different approach where we use two separate models. First a binary classification problem is solved and this decides, for each level in the well, whether it is fully water-saturated, or whether it contains some amount of hydrocarbons. For each level that is classified as water-only, the water saturation is simply set at one (100\%) and hydrocarbon saturation to zero. For each level that the classifier predicts contains some hydrocarbons, we run a regression model to predict the amount of water at that depth and set the hydrocarbon amount as the inverse of this accordingly. The idea of this approach is that the parts of the well that are water-only, which tends to be the vast majority of levels, will have their predicted saturation value set to precisely one which avoids noise in the predictions and false positives for the hydrocarbons.

\begin{figure}
	\begin{center}
		\includegraphics[scale=0.40]{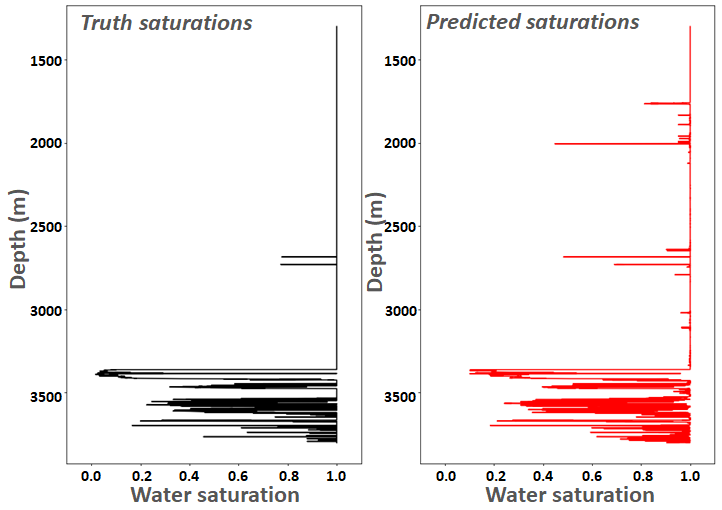}
	\end{center}
	\caption{Hydrocarbons saturation by depth using two models, boosted trees classification and regression, manually interpreted (truth) saturation in the plot on the left (black curve) and our prediction in the plot on the right (red curve)}
	\label{fig-boosted-both-hc-saturation-by-depth}
\end{figure}

\begin{figure}
	\begin{center}
		\includegraphics[scale=0.40]{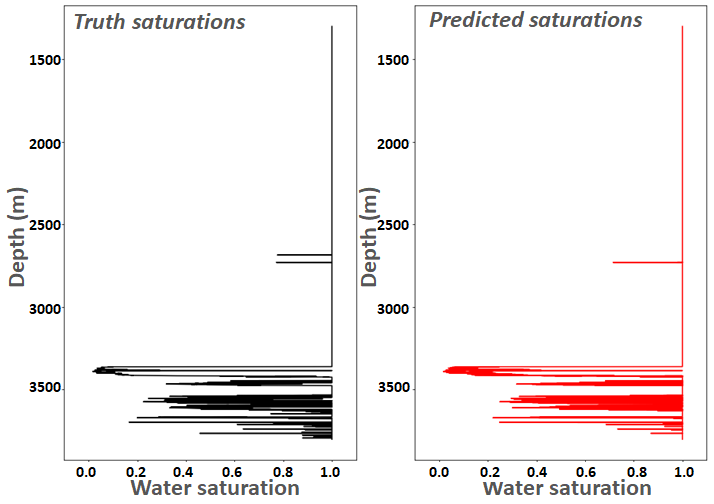}
	\end{center}
	\caption{Hydrocarbons saturation by depth using two models, Deep Neural Network (DNN) for classification and regression, manually interpreted (truth) saturation in the plot on the left (black curve) and our prediction in the plot on the right (red curve)}
	\label{fig-nn-both-hc-saturation-by-depth}
\end{figure}

Figure \ref{fig-boosted-both-hc-saturation-by-depth} illustrates the prediction of water saturation using boosted trees for both the classification and regression models. It can be seen that this really does not help and the accuracy of predictions, in comparison to the water predictions of Figure \ref{fig-simple-water-saturation-by-depth}, are way off. This is because the boosted trees classifier is seeing very many false positives of levels which it thinks contain some amount of hydrocarbons. These levels are then fed into the regression, and because it knows that the value will definitely not be 100\% water, values are predicted and errors introduced. We have a further option here because, unlike the vast majority of the well, these reservoirs tend to be very well covered by the measurements and-so missing input curves are rare. As such we can experiment with using a Deep Neural Network (DNN) instead of boosted trees to explore whether this can provide any improvement in accuracy. Figure \ref{fig-nn-both-hc-saturation-by-depth} illustrates the same experiment but where we use a DNN for both the classification and regression. It can be seen that, unlike the boosted trees prediction which under-predicts water, the DNN is over-predicting water and missing situations where hydrocarbons are present. Whilst the DNN is more conservative than boosted trees in predicting hydrocarbons, this conservatism is far more extreme in the DNN regression model, and there are a number of situations where the DNN classifier is predicting the presence of hydrocarbons but the DNN regression model then incorrectly predicts these to be a tiny amount. 

\begin{figure}
	\begin{center}
		\includegraphics[scale=0.40]{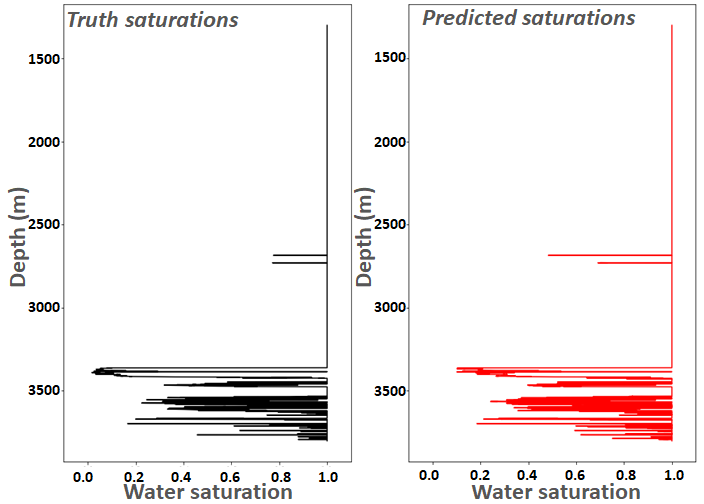}
	\end{center}
	\caption{Hydrocarbons saturation by depth using two models, deep neural network for classification and boosted trees for regression, manually interpreted (truth) saturation in the plot on the left (black curve) and our prediction in the plot on the right (red curve)}
	\label{fig-nnclass-boostedregress-hc-saturation-by-depth}
\end{figure}

So in fact we have a situation where the DNN classifier is more accurate than boosted trees classifier, and the boosted trees regression is more accurate than the DNN regression. Therefore, we decided to combine the best of both approaches, using our DNN for classification and boosted trees model for regression. Results from water saturation predictions using this hybrid DNN classification, boosted trees regression approach are illustrated in Figure \ref{fig-nnclass-boostedregress-hc-saturation-by-depth}. As can be seen this combines the best of both worlds, and whilst the accuracy of our fluid saturations predictions are not quite as good as those of our porosity predictions, this is the most accurate configuration we have found, they are still fairly accurate, and within accuracy limits set by the petrophysicists. Additionally, because of the nature of our boosted trees regression, this approach tends to favour over predicting hydrocarbons. This is useful as it is more important for the petrophysicists to have false positive for oil or gas that they can then explore and discount, rather than missing these areas altogether. Whilst our fluid saturations still require some form of human interpretation, analysis and validation, this provides them with a strong starting point.

\section{Lithoclass determination}
\label{sec:lithoclass}
In this last stage of the petrophysical workflow, lithology class or faces, which is the general geological rock type, is determined. This is a classification problem and categorises the rock as one of a number of different types. Out of the entirety of the workflow this is the simplest stage and as described in Section \ref{sec:related}, numerous other supervised learning studies have looked at this in detail. To some extent the problem of using supervised learning for lithology prediction has been largely solved. Therefore, instead we decided to investigate whether a more general approach could be used where we start from data without explicit lithoclass labels, apply some general lithoclass categorisation rules which have been provided by the petrophysicists, which effectively labels the data, and then train our model on this data. This is not unsupervised learning, which is where inferences are drawn from data-sets that have no labelled data to train on, because we are explicitly labelling some unlabelled data based on some generic rules and seeing how accurately we can make predictions based on this. But it is a useful technique to explore because often the well log data does not have lithology explicitly labelled, like other work in Section \ref{sec:related} assumes. Therefore a question is whether we can make accurate predictions from models trained on a set of generic membership rules which are based on mineralogy composition.

For this stage we are using the k-nearest neighbours algorithm for classification and the output of this algorithm is an item's class membership based on the properties of its neighbours, with each item being assigned to the class most common among its k nearest neighbours. We use the same six input curves, with cleaned p-wave and density curves, along with mineralogy and are focused on classifying records as hydrocarbon (HC) sand, shale, shaly sand, and wet sand  using relationships provided to use by the petrophysicists.

\begin{figure}
	\begin{center}
		\includegraphics[scale=0.60]{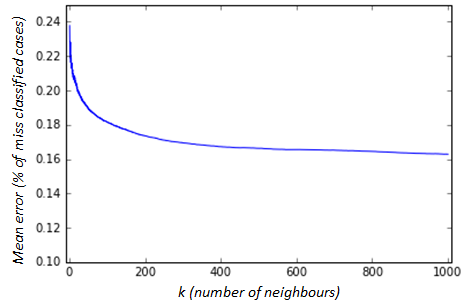}
	\end{center}
	\caption{The mean classification error (percentage of miss-classified cases) against k, the number of nearest neighbours to use in the classification.}
	\label{fig-lithology-k}
\end{figure}

A key configuration with k-nearest neighbours is what value of \emph{k} to use, i.e. the number of closest neighbours to each point that need to be considered in the classification. Using the K nearest neighbours classifier from Sklearn, to build our classification prediction, we then use a cross-validation approach to randomly pick samples and check that the classification is correct. Figure \ref{fig-lithology-k} illustrates how the mean classification error, which is the percentage of miss classified cases, relates to the number of neighbours to use in the classification (the value of \emph{k}). From these results we can see that the miss classification error reduces until between 200 and 300 neighbours, and at this point appears to level off. Therefore, for the lithology classification we choose \emph{k} to be 300 and train the classifier on the whole training set.

Figure \ref{fig-lithology-prediction} illustrates the number of rows in our test wells that have predicted with a specific lithology (columns) against the true lithology (rows), where we ideally want the maximum value for each prediction in the corresponding truth cell. Bearing in mind these predictions are based on general membership rules, the predictions for hydrocarbon sand, shale and wet sand are fairly reliable. However the model struggles with shaly sand. Whilst the accuracy of these predictions is lower than those of \cite{lithologyxgboost}, they are using data where the lithology is explicitly labelled by the petrophysicists to train their models. Instead we are applying a set of very simple rules and are able to generate predictions which are reasonable. This is interesting as it can be very quickly applied to large, unlabelled, data sets and then used for training. In such cases, apart from shaly sand, the petrophysicist can have a reasonable confidence in their predictions as a first step.

\begin{figure}
\begin{center}
\begin{tabular}{ | c c c c c | }
\hline
 Facies & HC Sand & Shale & Shaly Sand & Wet Sand \\ \hline
 HC Sand & 673 & 8 & 0 & 13 \\
 Shale & 2 & 27833 & 2259 & 653 \\
 Shaly Sand & 4 & 1735 & 2431 & 1228 \\
 Wet Sand & 18 & 306 & 2313 & 9369 \\
 \hline
\end{tabular}
\end{center}
\caption{Lithology prediction, number of rows in the test wells with a predicted and/or truth value. The columns are the predicted lithology and rows are the true lithology}
\label{fig-lithology-prediction}
\end{figure}

\section{Hyper-parameter search parallelisation}
\label{sec:hyperparams}
As described briefly in Section \ref{sec:cleaning} there are seven hyper-parameters that we must set for our boosted trees models. These are summarised in Figure \ref{fig-boosted-parameters} and they are interconnected, such that modifying the value of one parameter will impact the most suitable value of other parameters. This is especially challenging when it comes to boosted trees because they are sensitive to these hyper-parameters \cite{boosted-sensitive}, where there is a fairly small window of optimal hyper-parameters and outside this either the model under or over fits. The big challenge is that it is not clear what the correct hyper-parameter settings should be, nor how far from optimal they are. As such throughout this work we used Hyperopt \cite{hyperopt}, a Python library for automatically searching the hyper-parameter space and making optimal choices. Providing both a random search and tree of Parzen estimators \cite{parzen}, the user provides a description of their hyper-parameters, including the range of appropriate values. An objective function is also provided which returns a user defined \emph{loss} value, which is effectively what the framework aims to minimise.

\begin{figure}
\begin{center}
\begin{tabular}{ | c | c | }
\hline
 Name & Description \\ \hline
 colsample\_bytree & \makecell{Sub sample ratio of columns \\ when constructing each tree} \\ \hline 
 eta & \makecell{Step size shrinkage, \\ to prevent over-fitting}\\\hline
 gamma & \makecell{Minimum loss reduction required \\ to further partition a node}\\\hline
 max\_depth & \makecell{Maximum tree depth, the deeper  \\the tree the more complex \\ the model and likely to overfit}\\\hline
 min\_child\_weight & \makecell{Minimum sum of instance \\ weight needed in a child}\\\hline
 num\_rounds & \makecell{Number of boosting \\ rounds to perform}\\\hline
 subsample & \makecell{Sub sample ratio of the training \\ instances, useful to prevent over-fitting}\\
 \hline
\end{tabular}
\end{center}
\caption{Applicable boosted trees hyper-parameters and their description}
\label{fig-boosted-parameters}
\end{figure}

Therefore in this paper we have not explicitly mentioned the settings of these 7 hyper-parameters for each of our boosted trees models, because we ran ensembles of boosted trees models when training, relying on Hyperopt to search the parameter space and identify the most appropriate hyper-parameter settings for us. Hence when we talk about \emph{training a boosted trees model} in this paper, we implicitly mean performing this ensemble run of many individual boosted trees models and hyper-parameter optimisation because it is so important. Hyper-parameter searching can take a long time, especially because the code is serial when it comes to HPC machines such as the Cray XC30 we used in this work. Whilst the Hyperopt developers do claim to have a distributed version of the framework, crucially it is distributed via AWS YARN or Spark,  neither of which are available on the Cray we are using. We found on average it required between 120 and 160 hyper-parameters settings to be searched before we could be confident that our hyper-parameters were a good match to the model. Bearing in mind that training a single boosted trees model takes between ten and fifteen minutes then we are looking at around 20 hours in the best case and over double that in the worst case to train the models that we have used in this paper. The optimal hyper-parameters change not only on a model by model basis, e.g. a model that is trained for predicting Clay will require very different hyper-parameters to one predicting Quartz or Calcite, but also whenever we experimented with aspects such as the number or type of input curves. 

\begin{figure}
	\begin{center}
		\includegraphics[scale=0.70]{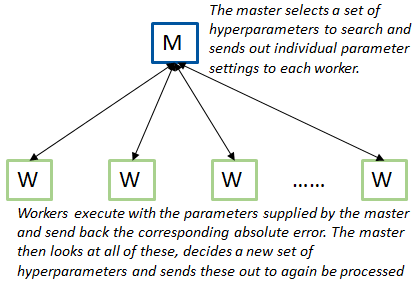}
	\end{center}
	\caption{Parallelisation of Hyperopt using master-worker pattern}
	\label{fig-hyperopt}
\end{figure}

In order to address this issue we developed an MPI implementation of the Hyperopt distribution layer. Using the MPI4Py Python library \cite{mpi4py}, we used the master-worker pattern as illustrated in Figure \ref{fig-hyperopt}, to distribute the searching of hyper-parameters across the nodes of the Cray. The master and each worker is a separate MPI process, and the master starts off by generating initial parameters settings for each worker to use as the settings when training their boosted trees model concurrently (one worker per NUMA region, as XGBoost uses OpenMP to parallelise across threads in this NUMA region). As workers feed back their resultant loss values the master will use this to then influence existing and further parameter choices which are then sent out to idle workers as they become available.

This is a very simple parallelisation strategy and, partly due to the maturity of MPI4Py took less than an hour to implement. However we found this ability to distribute over the nodes of ARCHER, the Cray XC30 used for this work, very useful when it came to productivity and taking full advantage of the XC30. From a parallelism perspective this design is fairly embarrassingly parallel, with the only communications between the master and workers needed at the start of each model iteration to communicate the hyper-parameter settings and at the end to send the resulting loss value back. Hence this scales well and typically we run over twenty nodes (480 cores), with two workers (boosted trees models) per node (as there are two NUMA regions per node in ARCHER.) This reduced the overall training runtime of our boosted trees models, including hyper-parameter optimisation, down from between 20 to 40 hours, to between 40 minutes and an hour. This was important because it resulted in a very significant increase in productivity.

\section{Conclusion}
\label{sec:conclusions}
In this paper we have studied the role that machine learning can play in tackling the entire petrophysical workflow for conditioning well log data. This is the first time that machine learning has been applied to the entire workflow and we have demonstrated reasonable prediction capabilities across the variety of workflow activities. Whilst using machine learning for some of the petrophysical activities, such as the cleaning of p-wave and density curves, along with the prediction of porosity and fluid saturations is highly accurate, it does struggle more with other aspects such as the mineralogy composition. Generally speaking this was not entirely unexpected, as the petrophysicists rely on more intuition from sources external to the well log data for mineralogy in comparison to the other stages. Whilst undoubtedly some of their knowledge and experience can be taught to a mathematical model by machine learning, our mineralogy predictions illustrate that there are limitations to this. These are important, novel, insights, both in terms of successes such as the combination of DNN classification and boosted trees regression providing accurate fluid saturations, and also the limitations of machine learning in this context such as the fact that geological formations doesn't really help improve mineralogy predictions.

In terms of the petrophysical interpretation time, we have not quite gone down from 7 days to 7 minutes, but once trained our models do very quickly, in a matter of seconds, provide predictions that can take humans many hours to equal. Whilst it is clear that machine learning is not going to replace the petrophysicists with such trained models any time soon, we do believe that machine learning has an important role to play in petrophysical interpretation and the use of this technique will continue to grow rapidly in the oil and gas industry. RSI petrophysicists have identified, based upon this research, two general benefits that machine learning provides here. Firstly as an initial, but very important, step in the interpretation where the use of the human is optimised by our models performing much of the time consuming mundane work. The idea being that the experienced petrophysicst is then presented with an estimation of mineralogy composition, porosity, fluid saturation, and lithology and from this can then tune and tweak the predictions to make them more robust. The second application of machine learning that has been identified from this work is as a quick, rough and ready pass, to determine whether a specific well is likely to contain features of interest (i.e. oil or gas), and warrant an in-depth manual interpretation or not. This fits in with a common industrial use-case, where the large oil and gas companies will provide geological experts at companies such as RSI with a variety of wells and up until this point there is little option but to perform a full interpretation. The ability to quickly and cheaply prioritise the most interesting wells is an important capability which machine learning provides.

We believe that it is a very exciting time for machine learning in the oil and gas industry, and there is plenty of further work that follows on from this study that will not only improve the accuracy of predictions but also apply machine learning to the wider area of sub-surface data analysis. From the mineralogy it is clear that the inclusion of mudlog data would be useful to provide additional context and improve prediction accuracy. The inclusion of this handwritten mudlog information will increase the complexity significantly, with advanced data extraction and pre-processing needing to be performed. But in conjunction with our existing well log mineralogy model we believe that there is significant potential here and this work will act as a baseline to understand the improved prediction accuracy that this affords. 

It is clear from our mineralogy experimentation that, given complete data, there is potential to improve the prediction accuracy using DNNs. One option here might be to use boosted trees as a first pass to estimate missing values and then feed these estimates into a DNN model. Even if the estimated value used is still fairly rough, this might be enough to gain good predictions with the DNN model. 

When parallelising Hyeropt with MPI it surprised us at how much low hanging fruit there is when it comes to these machine learning frameworks running on HPC machines. The fact that we were able, with a trivial amount of effort, to increase our productivity so significantly, illustrates the role that the HPC community and their expertise has to play in the engineering of these machine learning frameworks and enabling them to take advantage of large scale distributed machines. 


\section*{Acknowledgments}
The authors would like to thank the Oil and Gas Innovation Centre for funding this work and Rock Solid Images for access to their data and support in this project. We would also like to thank EPCC for providing us with time on ARCHER to train and test our models.

\end{document}